# Weighted Sets of Probabilities and Minimax Weighted Expected Regret: New Approaches for Representing Uncertainty and Making Decisions[*]

**Joseph Y. Halpern**
Cornell University
halpern@cs.cornell.edu

**Samantha Leung**
Cornell University
samlyy@cs.cornell.edu

## Abstract

We consider a setting where an agent's uncertainty is represented by a set of probability measures, rather than a single measure. Measure-by-measure updating of such a set of measures upon acquiring new information is well-known to suffer from problems; agents are not always able to learn appropriately. To deal with these problems, we propose using *weighted sets of probabilities*: a representation where each measure is associated with a *weight*, which denotes its significance. We describe a natural approach to updating in such a situation and a natural approach to determining the weights. We then show how this representation can be used in decision-making, by modifying a standard approach to decision making—minimizing expected regret—to obtain *minimax weighted expected regret* (MWER). We provide an axiomatization that characterizes preferences induced by MWER both in the static and dynamic case.

## 1 Introduction

Agents must constantly make decisions; these decisions are typically made in a setting with uncertainty. For decisions based on the outcome of the toss of a fair coin, the uncertainty can be well characterized by probability. However, what is the probability of you getting cancer if you eat fries at every meal? What if you have salads instead? Even experts would not agree on a single probability.

Representing uncertainty by a single probability measure and making decisions by maximizing expected utility leads to further problems. Consider the following stylized problem, which serves as a running example in this paper. The baker's delivery robot, T-800, is delivering $1,000$ cupcakes from the bakery to a banquet. Along the way, T-800 takes a tumble down a flight of stairs and breaks some of the cupcakes. The robot's map indicates that this flight of stairs must be either ten feet or fifteen feet high. For simplicity, assume that a fall of ten feet results in one broken cupcake, while a fall of fifteen feet results in ten broken cupcakes.

|  | 1 broken | 10 broken |
|---|---|---|
| *cont* | 10,000 | -10,000 |
| *back* | 0 | 0 |
| *check* | 5,001 | -4,999 |

Table 1: Payoffs for the robot delivery problem. Acts are in the leftmost column. The remaining two columns describe the outcome for the two sets of states that matter.

T-800's choices and their consequences are summarized in Table 1. Decision theorists typically model decision problems with states, acts, and outcomes: the world is in one of many possible states, and the decision maker chooses an *act*, a function mapping states to outcomes. A natural state space in this problem is $\{good, broken\}^{1000}$, where each state is a possible state of the cupcakes. However, all that matters about the state is the number of broken cakes, so we can further restrict to states with either one or ten broken cakes.

T-800 can choose among three acts: *cont*: continue the delivery attempt; *back*: go back for new cupcakes; or *check*: open the container and count the number of broken cupcakes, and then decide to continue or go back, depending on the number of broken cupcakes. The client will tolerate one broken cupcake, but not ten broken cupcakes. Therefore, if T-800 chooses *cont*, it obtains a utility of $10,000$ if there is only one broken cake, but a utility of $-10,000$ if there are ten broken cakes. If T-800 chooses to go *back*, then it gets a utility of $0$. Finally, checking the cupcakes costs $4,999$ units of utility but is reliable, so if T-800 chooses *check*, it ends up with a utility of $5,001$ if there is one broken cake, and a utility of $-4,999$ if there are ten broken cakes.

[*] Work supported in part by NSF grants IIS-0534064, IIS-0812045, and IIS-0911036, by AFOSR grants FA9550-08-1-0438 and FA9550-09-1-0266, and by ARO grant W911NF-09-1-0281.

If we try to maximize expected utility, we must assume some probability over states. What measure should be used? There are two hypotheses that T-800 entertains: (1) the stairs are ten feet high and (2) the stairs are fifteen feet high. Each of these places a different probability on states. If the stairs are ten feet high, we can take all of the $1,000$ states where there is exactly one broken cake to be equally probable, and take the remaining states to have probability 0; if the stairs are fifteen feet high, we can take all of the $C(1000, 10)$ states where there are exactly ten broken cakes to be equally probable, and take the remaining states to have probability 0. One way to model T-800's uncertainty about the height of the stairs is to take each hypothesis to be equally likely. However, not having any idea about which hypothesis holds is very different from believing that all hypotheses are equally likely. It is easy to check that taking each hypothesis to be equally likely makes *check* the act that maximizes utility, but taking the probability that the stairs are fifteen feet high to be .51 makes *back* the act that maximizes expected utility, and taking the probability that the stairs are ten feet high to be .51 makes *cont* the act that maximizes expected utility. What makes any of these choices the "right" choice?

It is easy to construct many other examples where a single probability measure does not capture uncertainty, and does not result in what seem to be reasonable decisions, when combined with expected utility maximization. A natural alternative, which has often been considered in the literature, is to represent the agent's uncertainty by a *set* of probability measures. For example, in the delivery problem, the agent's beliefs could be represented by two probability measures, $\Pr_1$ and $\Pr_{10}$, one for each hypothesis. Thus, $\Pr_1$ assigns uniform probability to all states with exactly one broken cake, and $\Pr_{10}$ assigns uniform probability to all states with exactly ten broken cakes.

But this representation also has problems. Consider the delivery example again. In a more realistic scenario, why should T-800 be sure that there is exactly either one broken cake or ten broken cakes? Of course, we can replace these two hypotheses by hypotheses that say that the probability of a cake being broken is either .001 or .01, but this doesn't solve the problem. Why should the agent be sure that the probability is either exactly .001 or exactly .01? Couldn't it also be .0999? Representing uncertainty by a set of measures still places a sharp boundary on what measures are considered possible and impossible.

A second problem involves updating beliefs. How should beliefs be updated if they are represented by a set of probability measures? The standard approach for updating a single measure is by conditioning. The natural extension of conditioning to sets of measure is measure-by-measure updating: conditioning each measure on the information (and also removing measures that give the information probability 0).

However, measure-by-measure updating can produce some rather counterintuitive outcomes. In the delivery example, suppose that a passer-by tells T-800 the information $E$: the first 100 cupcakes are good. Assuming that the passer-by told the truth, intuition tells us that there is now more reason to believe that there is only one broken cupcake.

However, $\Pr_1 \mid E$ places uniform probability on all states where the first 100 cakes are good, and there is exactly one broken cake among the last 900. Similarly, $\Pr_{10} \mid E$ places uniform probability on all states where the first 100 cakes are good, and there are exactly ten broken cakes among the last 900. $\Pr_1 \mid E$ still places probability 1 on there being one broken cake, just like $\Pr_1$, $\Pr_{10} \mid E$ still places probability 1 on there being ten broken cakes. There is no way to capture the fact that T-800 now views the hypothesis $\Pr_{10}$ as less likely, even if the passer-by had said instead that the first 990 cakes are all good!

Of course, both of these problems would be alleviated if we placed a probability on hypotheses, but, as we have already observed, this leads to other problems. In this paper, we propose an intermediate approach: representing uncertainty using *weighted sets of probabilities*. That is, each probability measure is associated with a weight. These weights can be viewed as probabilities; indeed, if the set of probabilities is finite, we can normalize them so that they are effectively probabilities. Moreover, in one important setting, we update them in the same way that we would update probabilities, using likelihood (see below). On the other hand, these weights do not act like probabilities if the set of probabilities is infinite. For example, if we had a countable set of hypotheses, we could assign them all weight 1 (so that, intuitively, they are all viewed as equally likely), but there is no uniform measure on a countable set.

More importantly, when it comes to decision making, we use the weights quite differently from how we would use second-order probabilities on probabilities. Second-order probabilities would let us define a probability on events (by taking expectation) and maximize expected utility, in the usual way. Using the weights, we instead define a novel decision rule, *minimax weighted expected regret (MWER)*, that has some rather nice properties, which we believe will make it widely applicable in practice. If all the weights are 1, then MWER is just the standard *minimax expected regret (MER)* rule (described below). If the set of probabilities is a singleton, then MWER agrees with (subjective) expected utility maximization (SEU). More interestingly perhaps, if the weighted set of measures converges to a single measure (which will happen in one important special case, discussed below), MWER converges to SEU. Thus, the weights give us a smooth, natural way of interpolating between MER and SEU.

In summary, weighted sets of probabilities allow us to represent ambiguity (uncertainty about the correct probability distribution). Real individuals are sensitive to this ambiguity when making decisions, and the MWER decision rule takes this into account. Updating the weighted sets of probabilities using likelihood allows the initial ambiguity to be resolved as more information about the true distribution is obtained.

We now briefly explain MWER, by first discussing MER. MER is a probabilistic variant of the minimax regret decision rule proposed by Niehans [10] and Savage [13]. Most likely, at some point, we've second-guessed ourselves and thought "had I known this, I would have done that instead". That is, in hindsight, we regret not choosing the act that turned out to be optimal for the realized state, called the *ex post* optimal act. The *regret* of an act $a$ in a state $s$ is the difference (in utility) between the ex post optimal act in $s$ and $a$. Of course, typically one does not know the true state at the time of decision. Therefore the regret of an act is the worst-case regret, taken over all states. The *minimax regret* rule orders acts by their regret.

The definition of regret applies if there is no probability on states. If an agent's uncertainty is represented by a single probability measure, then we can compute the *expected regret* of an act $a$: just multiply the regret of an act $a$ at a state $s$ by the probability of $s$, and then sum. It is well known that the order on acts induced by minimizing expected regret is identical to that induced by maximizing expected utility (see [8] for a proof). If an agent's uncertainty is represented by a set $\mathcal{P}$ of probabilities, then we can compute the expected regret of an act $a$ with respect to each probability measure $\Pr \in \mathcal{P}$, and then take the worst-case expected regret. The MER (Minimax Expected Regret) rule orders acts according to their worst-case expected regret, preferring the act that minimizes the worst-case regret. If the set of measures is the set of *all* probability measures on states, then it is not hard to show that MER induces the same order on acts as (probability-free) minimax regret. Thus, MER generalizes both minimax regret (if $\mathcal{P}$ consists of all measures) and expected utility maximization (if $\mathcal{P}$ consists of a single measure).

MWER further generalizes MER. If we start with a *weighted* set of measures, then we can compute the weighted expected regret for each one (just multiply the expected regret with respect to $\Pr$ by the weight of $\Pr$) and compare acts by their worst-case weighted expected regret.

Sarver [12] also proves a representation theorem that involves putting a multiplicative weight on a regret quantity. However, his representation is fundamentally different from MWER. In his representation, regret is a factor only when comparing two *sets* of acts; the ranking of individual acts is given by expected utility maximization. By way of contrast, we do not compare sets of acts.

It is standard in decision theory to axiomatize a decision rule by means of a representation theorem. For example, Savage [14] showed that if an agent's preferences $\succeq$ satisfied several axioms, such as completeness and transitivity, then the agent is behaving as if she is maximizing expected utility with respect to some utility function and probabilistic belief.

If uncertainty is represented by a set of probability measures, then we can generalize expected utility maximization to *maxmin expected utility (MMEU)*. MMEU compares acts by their worst-case expected utility, taken over all measures. MMEU has been axiomatized by Gilboa and Schmeidler [7]. MER was axiomatized by Hayashi [8] and Stoye [16]. We provide an axiomatization of MWER. We make use of ideas introduced by Stoye [16] in his axiomatization of MER, but the extension seems quite nontrivial.

We also consider a dynamic setting, where beliefs are updated by new information. If observations are generated according to a probability measure that is stable over time, then, as we suggested above, there is a natural way of updating the weights given observations, using ideas of likelihood. The idea is straightforward. After receiving some information $E$, we update each probability $\Pr \in \mathcal{P}$ to $\Pr \mid E$, and take its weight to be $\alpha_{\Pr} = \Pr(E)/\sup_{\Pr' \in \mathcal{P}} \Pr'(E)$. Thus, the weight of $\Pr$ after observing $E$ is modified by taking into account the likelihood of observing $E$ assuming that $\Pr$ is the true probability. We refer to this method of updating weights as *likelihood updating*.

If observations are generated by a stable measure (e.g., we observe the outcomes of repeated flips of a biased coin) then, as the agent makes more and more observations, the weighted set of probabilities of the agent will, almost surely, look more and more like a single measure. The weight of the measures in $\mathcal{P}$ closest to the measure generating the observations will converge to 1, and the weight of all other measures will converge to 0. This would not be the case if uncertainty were represented by a set of probability measures and we did measure-by-measure updating, as is standard. As we mentioned above, this means that MWER will converge to SEU.

We provide an axiomatization for dynamic MWER with likelihood updating. We remark that a dynamic version of MMEU with measure-by-measure updating has been axiomatized by Jaffray [9], Pires [11], and Siniscalchi [15].

Likelihood updating is somewhat similar in spirit to an updating method implicitly proposed by Epstein and Schneider [5]. They also represented uncertainty by using (unweighted) sets of probability measures. They choose a threshold $\alpha$ with $0 < \alpha < 1$, update by conditioning, and eliminate all measures whose relative likelihood does not exceed the threshold. This approach also has the property that, over time, all that is left in $\mathcal{P}$ are the measures closest to the measure generating the observations; all other measures are eliminated. However, it has the drawback that it introduces a new, somewhat arbitrary, parameter $\alpha$.

Chateauneuf and Faro [2] also consider weighted sets of probabilities (they model the weights using what they call *confidence functions*), although they impose more constraints on the weights than we do. They then define and provide a representation of a generalization of MMEU using weighted sets of probabilities that parallels our generalization of MER. Chateauneuf and Faro do not discuss the dynamic situation; specifically, they do not consider how weights should be updated in the light of new information.

The rest of this paper is organized as follows. Section 2 introduces the weighted sets of probabilities representation, and Section 3 introduces the MWER decision rule. Axiomatic characterizations of static and dynamic MWER are provided in Sections 4 and 5, respectively. We conclude in Section 6. All proofs can be found in the full paper (http://cs.cornell.edu/home/halpern/papers/mwr.pdf).

## 2 Weighted Sets of Probabilities

A set $\mathcal{P}^+$ of *weighted probability measures* on a set $S$ consists of pairs $(\Pr, \alpha_{\Pr})$, where $\alpha_{\Pr} \in [0,1]$ and $\Pr$ is a probability measure on $S$.[1] Let $\mathcal{P} = \{\Pr : \exists \alpha (\Pr, \alpha) \in \mathcal{P}^+\}$. We assume that, for each $\Pr \in \mathcal{P}$, there is exactly one $\alpha$ such that $(\Pr, \alpha) \in \mathcal{P}^+$. We denote this number by $\alpha_{\Pr}$, and view it as the *weight of* $\Pr$. We further assume for convenience that weights have been normalized so that there is at least one measure $\Pr \in \mathcal{P}$ such that $\alpha_{\Pr} = 1$.[2] We remark that, just as we do, Chateaunef and Faro [2] take weights to be in the interval $[0,1]$. They impose additional requirements on the weights. For example, they require that the weight of a convex combination of two probability measures is at least as high as the weight of each one. This does not seem reasonable in our applications. For example, an agent may know that one of two measures is generating his observations, and give them both weight 1, while giving all other distributions weight 0.

As we observed in the introduction, one way of updating weighted sets of probabilities is by using likelihood updating. We use $\mathcal{P}^+ \mid E$ to denote the result of applying likelihood updating to $\mathcal{P}^+$. Define $\overline{\mathcal{P}}^+(E) = \sup\{\alpha_{\Pr} \Pr(E) : \Pr \in \mathcal{P}\}$; if $\overline{\mathcal{P}}^+(E) > 0$, set $\alpha_{\Pr|E} = \sup_{\{\Pr' \in \mathcal{P} : \Pr'|E = \Pr|E\}} \frac{\alpha_{\Pr'} \Pr'(E)}{\overline{\mathcal{P}}^+(E)}$. Note that given a measure $\Pr \in \mathcal{P}$, there may be several distinct measures $\Pr'$ in $\mathcal{P}$ such that $\Pr' \mid E = \Pr \mid E$. Thus, we take the weight of $\Pr \mid E$ to be the sup of the possible candidate values of $\alpha_{\Pr|E}$. By dividing by $\overline{\mathcal{P}}^+(E)$, we guarantee that $\alpha_{\Pr|E} \in [0,1]$, and that there is some measure $\Pr$ such that $\alpha_{\Pr|E} = 1$, as long as there is some pair $(\alpha_{\Pr}, \Pr) \in \mathcal{P}$ such that $\alpha_{\Pr} \Pr(E) = \overline{\mathcal{P}}^+(E)$. If $\overline{\mathcal{P}}^+(E) > 0$, we take $\mathcal{P}^+ \mid E$ to be

$$\{(\Pr \mid E, \alpha_{\Pr|E}) : \Pr \in \mathcal{P}\}.$$

If $\overline{\mathcal{P}}^+(E) = 0$, then $\mathcal{P}^+ \mid E$ is undefined.

In computing $\mathcal{P}^+ \mid E$, we update not just the probability measures in $\mathcal{P}$, but also their weights. The new weight combines the old weight with the likelihood. Clearly, if all measures in $\mathcal{P}$ assign the same probability to the event $E$, then likelihood updating and measure-by-measure updating coincide. This is not surprising, since such an observation $E$ does not give us information about the relative likelihood of measures. We stress that using likelihood updating is appropriate only if the measure generating the observations is assumed to be stable. For example, if observations of heads and tails are generated by coin tosses, and a coin of possibly different bias is tossed in each round, then likelihood updating would not be appropriate.

It is well known that, when conditioning on a single probability measure, the order that information is acquired is irrelevant; the same observation easily extends to sets of probability measures. As we now show, it can be further extended to weighted sets of probability measures.

**Proposition 1.** *Likelihood updating is consistent in the sense that for all $E_1, E_2 \subseteq S$, $(\mathcal{P}^+ \mid E_1) \mid E_2 = (\mathcal{P}^+ \mid E_2) \mid E_1 = \mathcal{P}^+ \mid (E_1 \cap E_2)$, provided that $\mathcal{P}^+ \mid (E_1 \cap E_2)$ is defined.*

## 3 MWER

We now define MWER formally. Given a set $S$ of states and a set $X$ of outcomes, an *act* $f$ (over $S$ and $X$) is a function mapping $S$ to $X$. For simplicity in this paper, we take $S$ to be finite. Associated with each outcome $x \in X$ is a utility: $u(x)$ is the utility of outcome $x$. We call a tuple $(S, X, u)$ a *(non-probabilistic) decision problem*. To define regret, we need to assume that we are also given a set $M \subseteq X^S$ of feasible acts, called the *menu*. The reason for the menu is that, as is well known (and we will demonstrate by example shortly), regret can depend on the menu. Moreover, we assume that every menu $M$ has utilities bounded from above. That is, we assume that for all menus $M$, $\sup_{g \in M} u(g(s))$ is finite. This ensures that the regret of each act is well defined.[3] For a menu $M$ and act

---

[1] In this paper, for ease of exposition, we take the state space $S$ to be finite, and assume that all sets are measurable. We can easily generalize to arbitrary measure spaces.

[2] While we could take weights to be probabilities, and normalize them so that they sum to 1, if $\mathcal{P}$ is finite, this runs into difficulties if we have an infinite number of measures in $\mathcal{P}$. For example, if we are tossing a coin, and $\mathcal{P}$ includes all probabilities on heads from $1/3$ to $2/3$, using a uniform probability, we would be forced to assign each individual probability measure a weight of 0, which would not work well in the definition of MWER.

[3] Stoye [17] assumes that, for each menu $M$, there is a finite set $A_M$ of acts such that $M$ consists of all the convex combinations of the acts in $A_M$. Our assumption is clearly much weaker than Stoye's.

$f \in M$, the regret of $f$ with respect to $M$ and decision problem $(S, X, u)$ in state $s$ is

$$reg_M(f, s) = \left(\max_{g \in M} u(g(s))\right) - u(f(s)).$$

That is, the regret of $f$ in state $s$ (relative to menu $M$) is the difference between $u(f(s))$ and the highest utility possible in state $s$ (among all the acts in $M$). The regret of $f$ with respect to $M$ and decision problem $(S, X, u)$ is the worst-case regret over all states:

$$\max_{s \in S} reg_M(f, s).$$

We denote this as $reg_M^{(S,X,u)}(f)$, and usually omit the superscript $(S, X, u)$ if it is clear from context. If there is a probability measure Pr over the states, then we can consider the *probabilistic decision problem* $(S, X, u, \text{Pr})$. The *expected regret* of $f$ with respect to $M$ is

$$reg_{M,\text{Pr}}(f) = \sum_{s \in S} \text{Pr}(s) reg_M(f, s).$$

If there is a set $\mathcal{P}$ of probability measures over the states, then we consider the $\mathcal{P}$-decision problem $(S, X, u, \mathcal{P})$. The maximum expected regret of $f \in M$ with respect to $M$ and $(S, X, u, \mathcal{P})$ is

$$reg_{M,\mathcal{P}}(f) = \sup_{\text{Pr} \in \mathcal{P}} \left(\sum_{s \in S} \text{Pr}(s) reg_M(f, s)\right).$$

Finally, if beliefs are modeled by weighted probabilities $\mathcal{P}^+$, then we consider the $\mathcal{P}^+$-decision problem $(S, X, u, \mathcal{P}^+)$. The maximum weighted expected regret of $f \in M$ with respect to $M$ and $(S, X, u, \mathcal{P}^+)$ is

$$reg_{M,\mathcal{P}^+}(f) = \sup_{\text{Pr} \in \mathcal{P}} \left(\alpha_{\text{Pr}} \sum_{s \in S} \text{Pr}(s) reg_M(f, s)\right).$$

The MER decision rule is thus defined for all $f, g \in X^S$ as

$$f \succeq_{M,\mathcal{P}}^{S,X,u} g \text{ iff } reg_{M,\mathcal{P}}^{(S,X,u)}(f) \leq reg_{M,\mathcal{P}}^{(S,X,u)}(g).$$

That is, $f$ is preferred to $g$ if the maximum expected regret of $f$ is less than that of $g$. We can similarly define $\succeq_{M,reg}$, $\succeq_{M,\text{Pr}}^{S,X,u}$, and $\succeq_{M,\mathcal{P}^+}^{S,X,u}$ by replacing $reg_{M,\mathcal{P}}^{(S,X,u)}$ by $reg_M^{(S,X,u)}$, $reg_{M,\text{Pr}}^{(S,X,u)}$, and $reg_{M,\mathcal{P}^+}^{(S,X,u)}$, respectively. Again, we usually omit the superscript $(S, X, u)$ and subscript Pr or $\mathcal{P}^+$, and just write $\succeq_M$, if it is clear from context.

To see how these definitions work, consider the delivery example from the introduction. There are $1,000$ states with one broken cake, and $C(1000, 10)$ states with ten broken cakes. The regret of each action in a state depends only on the number of broken cakes, and is given in Table 2. It is easy to see that the action that minimizes regret is *check*, with *cont* and *back* having equal regret. If we represent uncertainty using the two probability measures $\text{Pr}_1$ and $\text{Pr}_{10}$, the expected regret of each of the acts with respect to $\text{Pr}_1$ (resp., $\text{Pr}_{10}$) is just its regret with respect to states with one (resp. ten) broken cakes. Thus, the action that minimizes maximum expected regret is again *check*.

|  | 1 broken cake | | 10 broken cakes | |
| --- | --- | --- | --- | --- |
|  | Payoff | Regret | Payoff | Regret |
| *cont* | 10,000 | 0 | -10,000 | 10,000 |
| *back* | 0 | 10,000 | 0 | 0 |
| *check* | 5,001 | 4,999 | -4,999 | 4,999 |

Table 2: Payoffs and regrets for delivery example.

|  | 1 broken cake | | 10 broken cakes | |
| --- | --- | --- | --- | --- |
|  | Payoff | Regret | Payoff | Regret |
| *cont* | 10,000 | 10,000 | -10,000 | 10,000 |
| *back* | 0 | 20,000 | 0 | 0 |
| *check* | 5,001 | 14,999 | -4,999 | 4,999 |
| *new* | 20,000 | 0 | -20,000 | 20,000 |

Table 3: Payoffs and regrets for the delivery problem with a new choice added.

As we said above, the ranking of acts based on MER or MWER can change if the menu of possible choices changes. For example, suppose that we introduce a new choice in the delivery problem, whose gains and losses are twice those of *cont*, resulting in the payoffs and regrets described in Table 3. In this new setting, *cont* has a lower maximum expected regret $(10,000)$ than *check* $(14,999)$, so MER prefers *cont* over *check*. Thus, the introduction of a new choice can affect the relative order of acts according to MER (and MWER), even though other acts are preferred to the new choice. By way of contrast, the decision rules MMEU and SEU are *menu-independent*; the relative order of acts according to MMEU and SEU is not affected by the addition of new acts.

We next consider a dynamic situation, where the agent acquires information. Specifically, in the context of the delivery problem, suppose that T-800 learns $E$—the first 100 items are good. Initially, suppose that T-800 has no reason to believe that one hypothesis is more likely than the other, so assigns both hypotheses weight 1. Note that $P_1(E) = 0.9$ and $\text{Pr}_{10}(E) = C(900, 10)/C(1000, 10) \approx 0.35$. Thus, $\mathcal{P}^+ \mid E = \{(\text{Pr}_1 \mid E, 1), (\text{Pr}_{10} \mid E, C(900, 10)/(.9C(1000, 10))\}$.

We can also see from this example that MWER interpolates between MER and expected utility maximization. Suppose that a passer-by tells T-800 that the first $N$ cupcakes are good. If $N = 0$, MWER with initial weights 1 is the same as MER. On the other hand, if $N \geq 991$, then the likelihood of $\text{Pr}_{10}$ is 0, and the only measure that has effect is $\text{Pr}_1$, which means minimizing maximum weighted expected regret is just maximizing expected utility with respect to $\text{Pr}_1$. If $0 < N < 991$, then the likelihoods (hence weights)

of $\Pr_1$ and $\Pr_{10}$ are 1 and $\frac{C(1000-N,10)}{C(1000,10)} \times \frac{1000}{1000-N} < ((999-N)/999)^9$.[9] Thus, as $N$ increases, the weight of $\Pr_{10}$ goes to 0, while the weight of $\Pr_1$ stays at 1.

## 4 An axiomatic characterization of MWER

We now provide a representation theorem for MWER. That is, we provide a collection of properties (i.e., axioms) that hold of MWER such that a preference order on acts that satisfies these properties can be viewed as arising from MWER. To get such an axiomatic characterization, we restrict to what is known in the literature as the *Anscombe-Aumann* (AA) framework [1], where outcomes are restricted to lotteries. This framework is standard in the decision theory literature; axiomatic characterizations of SEU [1], MMEU [7], and MER [8, 16] have already been obtained in the AA framework. We draw on these results to obtain our axiomatization.

Given a set $Y$ (which we view as consisting of *prizes*), a *lottery* over $Y$ is just a probability with finite support on $Y$. Let $\Delta(Y)$ consist of all finite probabilities over $Y$. In the AA framework, the set of outcomes has the form $\Delta(Y)$. So now acts are functions from $S$ to $\Delta(Y)$. (Such acts are sometimes called *Anscombe-Aumann acts*.) We can think of a lottery as modeling objective uncertainty, while a probability on states models subjective uncertainty; thus, in the AA framework we have both objective and subjective uncertainty. The technical advantage of considering such a set of outcomes is that we can consider convex combinations of acts. If $f$ and $g$ are acts, define the act $\alpha f + (1-\alpha)g$ to be the act that maps a state $s$ to the lottery $\alpha f(s) + (1-\alpha)g(s)$.

In this setting, we assume that there is a utility function $U$ on prizes in $Y$. The utility of a lottery $l$ is just the expected utility of the prizes obtained, that is,

$$u(l) = \sum_{\{y \in Y : l(y) > 0\}} l(y) U(y).$$

This makes sense since $l(y)$ is the probability of getting prize $y$ if lottery $l$ is played. The expected utility of an act $f$ with respect to a probability $\Pr$ is then just $u(f) = \sum_{s \in S} \Pr(s) u(f(s))$, as usual. We also assume that there are at least two prizes $y_1$ and $y_2$ in $Y$, with different utilities $U(y_1)$ and $U(y_2)$.

Given a set $Y$ of prizes, a utility $U$ on prizes, a state space $S$, and a set $\mathcal{P}^+$ of weighted probabilities on $S$, we can define a family $\succeq_{M,\mathcal{P}^+}^{S,\Delta(Y),u}$ of preference orders on Anscombe-Aumann acts determined by weighted regret, one per menu $M$, as discussed above, where $u$ is the utility function on lotteries determined by $U$. For ease of exposition, we usually write $\succeq_{M,\mathcal{P}^+}^{S,Y,U}$ rather than $\succeq_{M,\mathcal{P}^+}^{S,\Delta(Y),u}$.

We state the axioms in a way that lets us clearly distinguish the axioms for SEU, MMEU, MER, and MWER. The axioms are universally quantified over acts $f$, $g$, and $h$, menus $M$ and $M'$, and $p \in (0,1)$. We assume that $f, g \in M$ when we write $f \succeq_M g$.[4] We use $l^*$ to denote a constant act that maps all states to $l$.

**Axiom 1.** *(Transitivity)* $f \succeq_M g \succeq_M h \Rightarrow f \succeq_M h$.

**Axiom 2.** *(Completeness)* $f \succeq_M g$ or $g \succeq_M f$.

**Axiom 3.** *(Nontriviality)* $f \succ_M g$ for some acts $f$ and $g$ and menu $M$.

**Axiom 4.** *(Monotonicity)* If $(f(s))^* \succeq_{\{(f(s))^*, (g(s))^*\}} (g(s))^*$ for all $s \in S$, then $f \succeq_M g$.

**Axiom 5.** *(Mixture Continuity)* If $f \succ_M g \succ_M h$, then there exists $q, r \in (0,1)$ such that

$$qf + (1-q)h \succ_{M \cup \{qf+(1-q)h\}} g$$
$$\text{and} \quad g \succ_{M \cup \{rf+(1-r)h\}} rf + (1-r)h.$$

Menu-independent versions of Axioms 1–5 are standard. Clearly (menu-independent versions of) Axioms 1, 2, 4, and 5 hold for MMEU, MER, and SEU; Axiom 3 is assumed in all the standard axiomatizations, and is used to get a unique representation.

**Axiom 6.** *(Ambiguity Aversion)*

$$f \sim_M g \Rightarrow pf + (1-p)g \succeq_{M \cup \{pf+(1-p)g\}} g.$$

Ambiguity Aversion says that the decision maker weakly prefers to hedge her bets. It also holds for MMEU, MER, and SEU, and is assumed in the axiomatizations for MMEU and MER. It is not assumed for the axiomatization of SEU, since it follows from the Independence axiom, discussed next. Independence also holds for MWER, provided that we are careful about the menus involved. Given a menu M and an act $h$, let $pM + (1-p)h$ be the menu $\{pf + (1-p)h : p \in M\}$.

**Axiom 7.** *(Independence)*

$$f \succeq_M g \text{ iff } pf + (1-p)h \succeq_{pM+(1-p)h} pg + (1-p)h.$$

Independence holds in a strong sense for SEU, since we can ignore the menus. The menu-independent version of Independence is easily seen to imply Ambiguity Aversion. Independence does not hold for MMEU.

Although we have menu independence for SEU and MMEU, we do not have it for MER or MWER. The following two axioms are weakened versions of menu independence that do hold for MER and MWER.

---
[4]Stoye [17] assumed that menus were convex, so that if $f, g \in M$, then so is $pf + (1-p)g$. We do not make this assumption, although our results would still hold if we did (with the axioms slightly modified to ensure that menus are convex). While it may seem reasonable to think that, if $f$ and $g$ are feasible for an agent, then so is $pf + (1-p)g$, this not always the case. For example, it may be difficult for the agent to randomize, or it may be infeasible for the agent to randomize with probability $p$ for some choices of $p$ (e.g., for $p$ irrational).

**Axiom 8.** *(Menu independence for constant acts) If $l^*$ and $(l')^*$ are constant acts, then $l^* \succeq_M (l')^*$ iff $l^* \succeq_{M'} (l')^*$.*

In light of this axiom, when comparing constant acts, we omit the menu.

An act $h$ is *never strictly optimal relative to $M$* if, for all states $s \in S$, there is some $f \in M$ such that $(f(s))^* \succeq (h(s))^*$.

**Axiom 9.** *(Independence of Never Strictly Optimal Alternatives (INA)) If every act in $M'$ is never strictly optimal relative to $M$, then $f \succeq_M g$ iff $f \succeq_{M \cup M'} g$.*

**Axiom 10.** *(Boundedness of menus) For every menu $M$, there exists a lottery $\bar{l} \in \Delta(Y)$ such that for all $f \in M$ and $s \in S$, $(f(s))^* \preceq \bar{l}^*$.*

The boundedness axiom enforces the assumption that we made earlier that every menu has utilities that are bounded from above. Recall that this assumption is necessary for regret to be finite.

**Theorem 1.** *For all $Y$, $U$, $S$, and $\mathcal{P}^+$, the family of preference orders $\succeq_{M,\mathcal{P}^+}^{S,Y,U}$ satisfies Axioms 1–10. Conversely, if a family of preference orders $\succeq_M$ on the acts in $\Delta(Y)^S$ satisfies Axioms 1–10, then there exist a weighted set $\mathcal{P}^+$ of probabilities on $S$ and a utility $U$ on $Y$ such that $\succeq_M = \succeq_{M,\mathcal{P}^+}^{S,Y,U}$. Moreover, $U$ is unique up to affine transformations, and $\mathcal{P}^+$ can be taken to be* maximal*, in the sense that if $\succeq_M = \succeq_{M,(\mathcal{P}')^+}^{S,Y,U}$, and $(\alpha, \Pr) \in (\mathcal{P}')^+$, then there exists $\alpha' \geq \alpha$ such that $(\alpha', \Pr) \in \mathcal{P}^+$.*

Showing that $\succeq_{M,\mathcal{P}^+}^{S,Y,U}$ satisfies Axioms 1–10 is fairly straightforward; we leave details to the reader. The proof of the converse is quite nontrivial, although it follows the lines of the proof of other representation theorems. We provide an outline of the proof here; details can be found in the full paper (http://cs.cornell.edu/home/halpern/papers/mwr.pdf).

Using standard techniques, we can show that the axioms guarantee the existence of a utility function $U$ on prizes that can be extended to lotteries in the obvious way, so that $l^* \succeq (l')^*$ iff $U(l) \geq U(l')$. We then use techniques of Stoye [17] to show that it suffices to get a representation theorem for a single menu, rather than all menus: the menu consisting of all acts $f$ such that $U(f(s)) \leq 0$ for all states $s \in S$. This allows us to use techniques in the spirit of those used by by Gilboa and Schmeidler [6] to represent (unweighted) MMEU. However, there are technical difficulties that arise from the fact that we do not have a key axiom that is satisfied by MMEU: C-independence (discussed below). The heart of the proof involves dealing with the lack of C-independence; We leave details of the proof to the full paper.

In standard representation theorems, not only is the utility function unique (up to affine transformations, so that we can replace $U$ by $aU + b$, where $a > 0$ and $b$ are constants), but the probability (or set of probabilities) is unique as well. We were not able to find natural conditions on weighted sets of probabilities that guarantee uniqueness. In the case of sets of probabilities, we need to assume that the set is convex and closed to get uniqueness. But there seems to be no natural notion of convexity for a set $\mathcal{P}^+$ of weighted probabilities, and the requirement that $\mathcal{P}^+$ be closed seems unreasonable. For example, if $\mathcal{P}^+$ consists of a single probability measure $\Pr$ with weight $\alpha_{\Pr} = 1$, then there are sequences $\Pr_n \to \Pr$ with $\alpha_{\Pr_n} = 0$, and yet $\alpha_{\Pr} = 1$. To see why something like convexity is needed for uniqueness, consider the delivery example and the expected regrets in Table 2, and the distribution $0.5\Pr_1 + 0.5\Pr_{10}$. The weighted expected regret of any act with respect to $0.5\Pr_1 + 0.5\Pr_{10}$ is bounded above by the maximum weighted expected regret of that act with respect to $\Pr_1$ and $\Pr_{10}$. Therefore, adding $0.5\Pr_1 + 0.5\Pr_{10}$ to $\mathcal{P}^+$, with any weight in $(0, 1]$, yields another representation for the MWER preferences. Although we do not get a unique set $\mathcal{P}^+$ in the representation, the maximality requirement allows us to view the set $\mathcal{P}^+$ as canonical in a certain sense.

It is instructive to compare Theorem 1 to other representation results in the literature. Anscombe and Aumann [1] showed that the menu-independent versions of axioms 1–5 and 7 characterize SEU. The presence of Axiom 7 (menu-independent Independence) greatly simplifies things. Gilboa and Schmeidler [7] showed that axioms 1–6 together with one more axiom that they call *Certainty-independence* characterizes MMEU. Certainty-independence, or C-independence for short, is a weakening of independence (which, as we observed, does not hold for MMEU), where the act $h$ is required to be a constant act. Since MMEU is menu-independent, we state it in a menu-independent way.

**Axiom 11.** *(C-Independence) If $h$ is a constant act, then $f \succeq g$ iff $pf + (1-p)h \succeq pg + (1-p)h$.*

As we observed, in general, we have Ambiguity Aversion (Axiom 6) for regret. *Betweenness* [3] is a stronger notion than ambiguity aversion, which states that if an agent is indifferent between two acts, then he must also be indifferent among all convex combinations of these acts. While betweenness does not hold for regret, Stoye [16] gives a weaker version that does hold. A menu $M$ has *state-independent outcome distributions* if the set $L(s) = \{y \in \Delta(Y) : \exists f \in M, f(s) = y\}$ is the same for all states $s$.

**Axiom 12.** *If $h$ is a constant act, and $M$ has state-independent outcome distributions, then*

$$h \sim_M f \Rightarrow pf + (1-p)h \sim_{M \cup \{pf+(1-p)h\}} f.$$

The assumption that the menu has state-independent outcome distributions is critical in Axiom 12.

Stoye [16] shows that Axioms 1–9 together with Axiom 12

|        | SEU | REG | MER | MWER | MMEU |
|--------|-----|-----|-----|------|------|
| Ax. 1-6,8-10 | ✓ | ✓ | ✓ | ✓ | ✓ |
| Ind    | ✓ | ✓ | ✓ | ✓ |   |
| C-Ind  | ✓ |   |   |   | ✓ |
| Ax. 12 | ✓ | ✓ | ✓ |   |   |
| Symmetry | ✓ | ✓ |   |   |   |

Table 4: Characterizing axioms for several decision rules.

characterize MER.[5] Non-probabilistic regret (which we denote REG) can be viewed as a special case of MER, where $\mathcal{P}$ consists of all distributions. This means that it satisfies all the axioms that MER satisfies. As Stoye [17] shows, REG is characterized by Axioms 1–9 and one additional axiom, which he calls Symmetry. We omit the details here.

The assumption that the menu has state-independent outcome distributions is critical in Axiom 12. In the full paper, we give an example showing that the variant of Axiom 12 without the state-independent outcome distribution requirement does not hold.

Table 4 describes the relationship between the axioms characterizing the decision rules.

## 5 Characterizing MWER with Likelihood Updating

We next consider a more dynamic setting, where agents learn information. For simplicity, we assume that the information is always a subset $E$ of the state space. If the agent is representing her uncertainty using a set $\mathcal{P}^+$ of weighted probability measures, then we would expect her to update $\mathcal{P}^+$ to some new set $\mathcal{Q}^+$ of weighted probability measures, and then apply MWER with uncertainty represented by $\mathcal{Q}^+$. In this section, we characterize what happens in the special case that the agent uses likelihood updating, so that $\mathcal{Q}^+ = (\mathcal{P}^+ \mid E)$.

For this characterization, we assume that the agent has a family of preference orders $\succeq_{E,M}$ indexed not just by the menu $M$, but by the information $E$. Each preference order $\succeq_{E,M}$ satisfies Axioms 1–10, since the agent makes decisions after learning $E$ using MWER. Somewhat surprisingly, all we need is one extra axiom for the characterization; we call this axiom MDC, for 'menu-dependent dynamic consistency'.

To explain the axiom, we need some notation. As usual, we take $fEh$ to be the act that agrees with $f$ on $E$ and with $h$ off of $E$; that is

$$fEh(s) = \begin{cases} f(s) & \text{if } s \in E \\ h(s) & \text{if } s \notin E. \end{cases}$$

---
[5]Stoye actually worked with choice correspondences; see Section 6.

In the delivery example, the act *check* can be thought of as $(cont)E(back)$, where $E$ is the set of states where there is only one broken cake.

Roughly speaking, MDC says that you prefer $f$ to $g$ once you learn $E$ if and only if, for any act $h$, you also prefer $fEh$ to $gEh$ before you learn anything. This seems reasonable, since learning that the true state was in $E$ is conceptually similar to knowing that none of your choices matter off of $E$.

To state MDC formally, we need to be careful about the menus involved. Let $MEh = \{fEh : f \in M\}$. We can identify unconditional preferences with preferences conditional on $S$; that is, we identify $\succeq_M$ with $\succeq_{S,M}$. We also need to restrict the sets $E$ to which MDC applies. Recall that conditioning using likelihood updating is undefined for an event such that $\overline{\mathcal{P}}^+(E) = 0$. That is, $\alpha_{\Pr} \Pr(E) = 0$ for all $\Pr \in \mathcal{P}$. As is commonly done, we capture the idea that conditioning on $E$ is possible using the notion of a *non-null* event.

**Definition 1.** *An event $E$ is* null *if, for all $f, g \in \Delta(Y)^S$ and menus $M$ with $fEg, g \in M$, we have $fEg \sim_M g$.*

**MDC.** *For all non-null events $E$, $f \succeq_{E,M} g$ iff $fEh \succeq_{MEh} gEh$ for some $h \in M$.*[6]

The key feature of MDC is that it allows us to reduce all the conditional preference orders $\succeq_{E,M}$ to the unconditional order $\succeq_M$, to which we can apply Theorem 1.

**Theorem 2.** *For all $Y$, $U$, $S$, and $\mathcal{P}^+$, the family of preference orders $\succeq_{M,\mathcal{P}^+|E}^{S,Y,U}$ for events $E$ such that $\overline{\mathcal{P}}^+(E) > 0$ satisfies Axioms 1–10 and MDC. Conversely, if a family of preference orders $\succeq_{E,M}$ on the acts in $\Delta(Y)^S$ satisfies Axioms 1–10 and MDC, then there exist a weighted set $\mathcal{P}^+$ of probabilities on $S$ and a utility $U$ on $Y$ non-null $E$, $\succeq_{E,M} = \succeq_{M,\mathcal{P}^+|E}^{S,Y,U}$. Moreover, $U$ is unique up to affine transformations, and $\mathcal{P}^+$ can be taken to be maximal.*

*Proof.* Since $\succeq_M = \succeq_{S,M}$ satisfies Axioms 1–10, there must exist a weighted set $\mathcal{P}^+$ of probabilities on $S$ and a utility function $U$ such that $f \succeq_M g$ iff $f \succeq_{M,\mathcal{P}^+}^{S,Y,U} g$. We now show that if $E$ is non-null, then $\overline{\mathcal{P}}^+(E) > 0$, and $f \succeq_{E,M} g$ iff $f \succeq_{M,\mathcal{P}^+|E}^{(S,X,u)} g$.

For the first part, it clearly is equivalent to show that if $\overline{\mathcal{P}}^+(E) = 0$, then $E$ is null. So suppose that $\overline{\mathcal{P}}^+(E) = 0$. Then $\alpha_{\Pr} \Pr(E) = 0$ for all $\Pr \in \mathcal{P}$. This means that $\alpha_{\Pr} \Pr(s) = 0$ for all $\Pr \in \mathcal{P}$ and $s \in E$. Thus, for all acts

---
[6]Although we do not need this fact, it is worth noting that the MWER decision rule has the property that $fEh \succeq_{MEh} gEh$ for some act $h$ iff $fEh \succeq_{MEh} gEh$ for all acts $h$. Thus, this property follows from Axioms 1–10.

$f$ and $g$,

$$\begin{aligned}
®_{M,\mathcal{P}^+}(fEg) \\
&= \sup\nolimits_{\Pr\in\mathcal{P}} \left(\alpha_{\Pr} \sum\nolimits_{s\in S} \Pr(s) reg_M(fEg, s)\right) \\
&= \sup\nolimits_{\Pr\in\mathcal{P}} \left(\alpha_{\Pr} \left(\sum\nolimits_{s\in E} \Pr(s) reg_M(f, s)\right) \right. \\
&\quad \left. + \sum\nolimits_{s\in E^c} \Pr(s) reg_M(g, s)\right) \\
&= \sup\nolimits_{\Pr\in\mathcal{P}} \left(\alpha_{\Pr} \sum\nolimits_{s\in S} \Pr(s) reg_M(g, s)\right) \\
&= reg_{M,\mathcal{P}^+}(g).
\end{aligned}$$

Thus, $fEg \sim_M g$ for all acts $f, g$ and menus $M$ containing $fEg$ and $g$, which means that $E$ is null.

For the second part, we first show that if $\overline{\mathcal{P}}^+(E) > 0$, then for all $f, h \in M$, we have that

$$reg_{MEh,\mathcal{P}^+}(fEh) = \overline{\mathcal{P}}^+(E) reg_{M,\mathcal{P}^+|E}(f).$$

We proceed as follows:

$$\begin{aligned}
®_{MEh,\mathcal{P}^+}(fEh) \\
&= \sup\nolimits_{\Pr\in\mathcal{P}} \left(\alpha_{\Pr} \sum\nolimits_{s\in S} \Pr(s) reg_{MEh}(fEH, s)\right) \\
&= \sup\nolimits_{\Pr\in\mathcal{P}} \left(\alpha_{\Pr} \Pr(E) \sum\nolimits_{s\in E} \Pr(s\mid E) reg_M(f, s) \right. \\
&\quad \left. + \alpha_{\Pr} \sum\nolimits_{s\in E^c} \Pr(s) reg_{\{h\}}(h, s)\right) \\
&= \sup\nolimits_{\Pr\in\mathcal{P}} \left(\alpha_{\Pr} \Pr(E) \sum\nolimits_{s\in E} \Pr(s|E) reg_M(s, f)\right) \\
&= \sup\nolimits_{\Pr\in\mathcal{P}} \left(\overline{\mathcal{P}}^+(E) \alpha_{\Pr|E} \sum\nolimits_{s\in E} \Pr(s|E) reg_M(f, s)\right) \\
&\quad [\text{since } \alpha_{\Pr|E} = \sup\nolimits_{\{\Pr'\in\mathcal{P}: \Pr'|E=\Pr|E\}} \frac{\alpha_{\Pr'}\Pr'(E)}{\overline{\mathcal{P}}^+(E)}] \\
&= \overline{\mathcal{P}}^+(E) \cdot reg_{M,\mathcal{P}^+|E}(f).
\end{aligned}$$

Thus, for all $h \in M$,

$$\begin{aligned}
®_{MEh,\mathcal{P}^+}(fEh) \leq reg_{MEh,\mathcal{P}^+}(gEh) \\
&\text{iff } \overline{\mathcal{P}}^+(E) \cdot reg_{M,\mathcal{P}^+|E}(f) \leq \overline{\mathcal{P}}^+(E) \cdot reg_{M,\mathcal{P}^+|E}(g) \\
&\text{iff } reg_{M,\mathcal{P}^+|E}(f) \leq reg_{M,\mathcal{P}^+|E}(g).
\end{aligned}$$

It follows that the order induced by $\Pr^+$ satisfies MDC.

Moreover, if 1–10 and MDC hold, then for the weighted set $\mathcal{P}^+$ that represents $\succeq_M$, we have

$$\begin{aligned}
&f \succeq_{E,M} g \\
&\text{iff } \text{for some } h \in M, fEh \succeq_{MEh} gEh \\
&\text{iff } reg_{M,\mathcal{P}^+|E}(f) \leq reg_{M,\mathcal{P}^+|E}(g),
\end{aligned}$$

as desired. $\square$

Analogues of MDC have appeared in the literature before in the context of updating preference orders. In particular, Epstein and Schneider [4] discuss a menu-independent version of MDC, although they do not characterize updating in their framework. Sinischalchi [15] also uses an analogue of MDC in his axiomatization of measure-by-measure updating of MMEU. Like us, he starts with an axiomatization for unconditional preferences, and adds an axiom called *constant-act dynamic consistency* (CDC), somewhat analogous to MDC, to extend the axiomatization of MMEU to deal with conditional preferences.

# 6 Conclusion

We proposed an alternative belief representation using *weighted sets of probabilities*, and described a natural approach to updating in such a situation and a natural approach to determining the weights. We also showed how weighted sets of probabilities can be combined with regret to obtain a decision rule, MWER, and provided an axiomatization that characterizes static and dynamic preferences induced by MWER.

We have considered preferences indexed by menus here. Stoye [17] used a different framework: *choice functions*. A choice function maps every finite set $M$ of acts to a subset $M'$ of $M$. Intuitively, the set $M'$ consists of the 'best' acts in $M$. Thus, a choice function gives less information than a preference order; it gives only the top elements of the preference order. The motivation for working with choice functions is that an agent can reveal his most preferred acts by choosing them when the menu is offered. In a menu-independent setting, the agent can reveal his whole preference order; to decide if $f \succ g$, it suffices to present the agent with a choice among $\{f, g\}$. However, with regret-based choices, the menu matters; the agent's most preferred choice(s) when presented with $\{f, g\}$ might no longer be the most preferred choice(s) when presented with a larger menu. Thus, a whole preference order is arguably not meaningful with regret-based choices. Stoye [17] provides a representation theorem for MER where the axioms are described in terms of choice functions. The axioms that we have attributed to Stoye are actually the menu-based analogue of his axioms. We believe that it should be possible to provide a characterization of MWER using choice functions, although we have not yet proved this.

Finally, we briefly considered the issue of dynamic consistency and consistent planning. As we showed, making this precise in the context of regret involves a number of subtleties. We hope to return to this issue in future work.

**Acknowledgments:** We thank Joerg Stoye and Edward Lui for useful comments.